\documentclass[a4paper,12pt]{article}
\usepackage[english]{babel}
\usepackage[utf8]{inputenc}
\usepackage{siunitx}
\usepackage{authblk}
\usepackage{glossaries}
\setacronymstyle{long-short}
\usepackage{cite}
\usepackage{hyperref}
\usepackage{cleveref}
\usepackage{latexsym}
\usepackage{amsmath}
\usepackage{multicol}
\usepackage[pagewise]{lineno}
\usepackage{xpatch}
\usepackage{lipsum}  

\def\enddisplaymath{\]\@ignoretrue}

\makeatletter         
\def\@maketitle{
\raggedleft
\vspace*{25px}

\begin{center}
{\Large \bfseries \sffamily \@title }\\[4ex] 
{\@author}\\[4ex] 
\@date\\[8ex]
\includegraphics[width = 55mm]{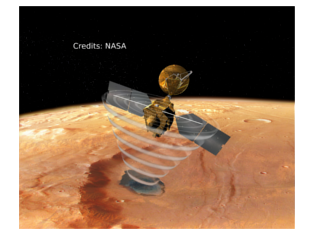}
\end{center}}
\makeatother

\usepackage{cleveref}
\usepackage{xcolor}
\usepackage[pagewise]{lineno}
\usepackage{fancyhdr}
\usepackage{lastpage}

\usepackage[ddmmyyyy]{datetime} 
\hypersetup{
    colorlinks=true,
    linkcolor=blue,
    filecolor=magenta,      
    urlcolor=cyan,
}
\newcounter{mybibstartvalue}
\xpatchcmd{\thebibliography}{%
  \usecounter{enumiv}%
}{%
  \usecounter{enumiv}%
  \setcounter{enumiv}{\value{mybibstartvalue}}%
}{}{}

\usepackage{graphicx}
\graphicspath{ {./pictures/} } 
\usepackage[font=small,labelfont=bf]{caption} 
\usepackage{booktabs} 
\usepackage{relsize} 
\DeclareUnicodeCharacter{00B0}{}
\usepackage{amsthm}
\theoremstyle{definition}
\fancypagestyle{plain}{%
  \fancyhf{}%
  \fancyhead[C]{\includegraphics[width = 15cm]{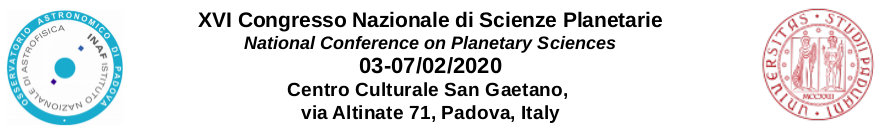} } %
}

\DeclareRobustCommand{\[}{\begin{equation*}}
\DeclareRobustCommand{\]}{\end{equation*}}

\title{ 
High Quality Software for\\ Planetary Science from Space}
\author[1]{Lazzarotto, F.}
\author[1]{Cremonese, G.}
\author[1]{Lucchetti, A.}
\author[1]{Re, C.}
\author[1]{Simioni, E.}
\author[1]{Pajola, M.}
\author[2,1]{Cambianica, P.}
\author[2,1]{Munaretto, G.}

\affil[1]{\small INAF-OAPD, Padova, Italy, \url{http://www.oapd.inaf.it}} 
\affil[2]{{\small University of Padova - DFA, Padova, Italy, \url{https://www.dfa.unipd.it}}}
\date{February 7, 2020 - {\small mailto:francesco.lazzarotto@inaf.it}}  
\newtheorem{mydef}{Definition}

\begin{document}

\maketitle

\begin{abstract}
Planetary science space missions need high quality software ed efficient algorithms 
in order to extract innovative scientific results from flight data.
Reliable and efficient software technologies are increasingly vital 
to improve and prolong the exploiting of the results of a mission, 
to allow the application of established algorithms and technologies 
also to future space missions and for the scientific analysis of archived data.
Here after will be given an in-depth analysis study accompanied by implementation examples 
on ESA and ASI missions and some remarkable results fruit of decades of important experience 
reached by space agencies and research institutes in the field. 
Space applications software quality analysis is not different from other application contexts, 
among the hi-tech and hi-reliability fields. 
We describe here a Software Quality study in general, then we will focus on the quality of 
space mission software (s/w) with details on some notable cases.
\end{abstract}

\section{Software Quality}
\mydef \emph{Quality: The degree to which a set of inherent characteristics of an entity fulfills requirements 
[ISO 9000:2005]}.\\
\includegraphics[width=1.0\linewidth]{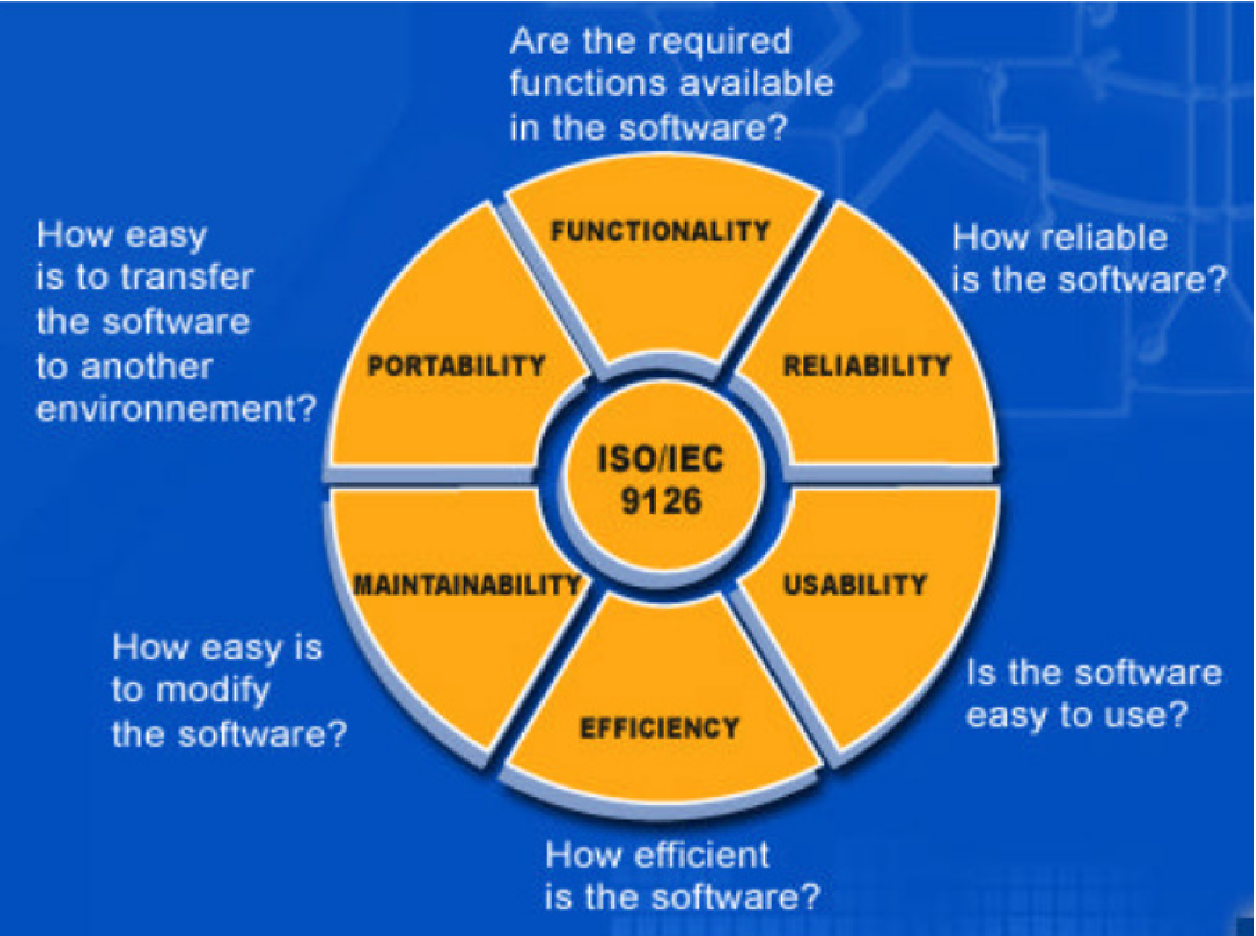}
\captionof{figure}{S/W quality attributes}
Quality means satisfying requirements which may be explicitly stated 
or implicitly given by a shared notion, it is not an absolute degree of goodness.
In \textbf{Software Quality}, the entity which shall fulfill requirements is a software system or component.
Evaluating the quality of software is not simple because usually s/w is not tangible and manifests 
only in effects and documentation, especially the source code.
We can face software quality studies with a pragmatic practical view or with the more formal ISO standards view.
\textbf{The pragmatic view}
\begin{itemize}
\item  External Quality (as perceived);
\item  Internal Quality (of s/w artifacts);
\end{itemize}
\textbf{The ISO standards’ view { \small [ISO/IEC 25010:2011] } } \\
based on measurement
\begin{itemize}
\item Internal measures: internal quality;
\item External measures: external quality;
\item Usage measures: quality in use.
\end{itemize}
\section{Software for Space Science}
There are not drawbacks for using widely adopted free standards 
and free and open source solutions for space missions data handling and computing software. 
This has been the key of success for the most important advances 
in general purpose computing and information technologies of last
decades (advent of www, cloud services, data science, smartphones, \ldots).
\begin{center}
\includegraphics[width=0.8\linewidth]{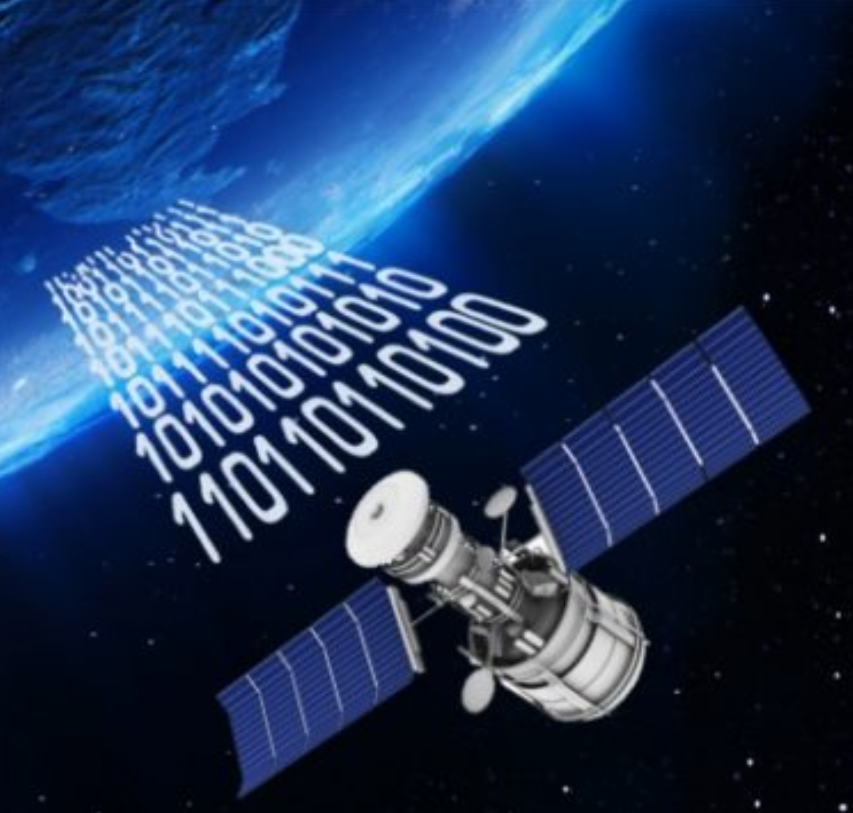}
\captionof{figure}{Satellite data transfer}
\end{center}
For (Astro)Physics and Planetary Science we can cite Astropy\cite{2018AJ....156..123A}, 
PlanetaryPy\cite{PlPy}, CERN/Root framework\cite{brun:96:aihenp}, GDAL by OSGeo\cite{gdalogr19}, 
QGis\cite{qgis19} and GNU Scientific Library\cite{gsl09}.
In our project we mostly used C/C++ and Bash computing languages, with some Python and Java modules,
from which was very easy and efficient to call the described libraries, with no drawbacks, extra-costs
and unnecessary network connections due to proprietary licensed software.
Pipelines commands can be also run step by step 
interactively or operated graphically through GUIs that 
has been simply and portably implemented (Win, Mac, Linux) using kdialog by KDE or via
more advanced compiled apps. 
All the software included in our s/w distributions has been saved under a version control system, 
such as Git, SVN or CVS to facilitate the team work, having reliable distributed access 
and to individuate official releases.


\section{BepiColombo mission}
BepiColombo\cite{BC} (BC) mission to Mercury by ESA/JAXA will provide measurements 
to allow studies of composition, origin and dynamics of Mercury’s exosphere 
and polar deposits as well as of the structure and dynamics of Mercury’s magnetosphere.
Expected to arrive at Mercury in 2025 after Earth and Venus flybys.
On 2018-10-20 03.45 (CEST), BC was successfully launched from 
european spaceport Kourou (French Guyana) with an Ariane 5 vehicle.
\begin{center}
\includegraphics[width=1.0\linewidth]{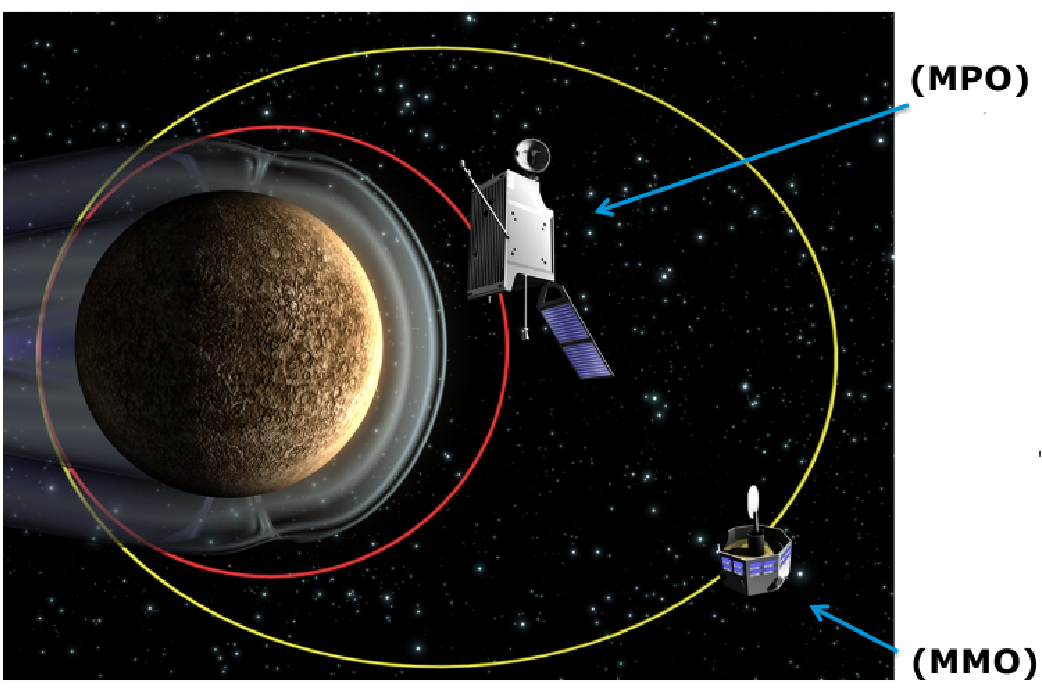}
\captionof{figure}{BC spacecrafts on Mercury orbits}
\end{center}

\section{Space Data Processing}
We created \textbf{SETM}: a set of s/w libraries to process satellite data, supporting different
aspects of BC space mission Data Handling and Operations tasks and we are applying it 
to the SERENA\cite{Orsini2019EPSC}\cite{2010P&SS...58..166O} and SIMBIO-SYS\cite{2016MmSAI..87..171F}\cite{2009aogs...15..305C} BC instruments. 
We focalised on widely adopted data structures and formats 
as well as we imported and implemented commonly used algorithms, being neutral from the specific 
formats and computing languages. 
The functions to be performed by the s/w system has been collected 
in a S/W Requirement Specification document (SRS), following ISO/IEC/IEEE 29148-2011 standard,
using instrument team requests and S/W engineering guidelines advised by ESA.\\ SETM functions include:
\begin{itemize}
\item TM decoding/encoding;
\item Data Structures \& Metadata definition;
\item Statistics, Plots \& Reporting; 
\item File I/O \& DB archiving;
\item Networking \& Automation.
\end{itemize}


\section{Conclusion}
The review of yet developed projects supported by ESA, EUMETSAT, NASA and NOAA 
and the experience on science data processing systems software 
for space missions participated by INAF led us 
to the synthesis of an efficient collection of s/w libraries 
developed with a general approach, containing all is needed to share and
reuse it in international collaborative research projects.
\section{References}
\renewcommand{\section}[2]{\vskip 0.05em} 
\nocite{*} 
{
\smaller
\bibliography{lazzarotto_poster_to_article_padovaplanet2020}{}
\bibliographystyle{plain}
}

\end{document}